\documentclass[twocolumn]{aastex62}
\usepackage{amsmath}
\usepackage{lineno}

\newcommand{\Msun}{\ensuremath{{\rm M}_\odot}}
\newcommand{\ledd}{\ensuremath{{\rm L}_{\rm Edd}}}
\graphicspath{{./}{figures/}}

\received{...}
\revised{...}
\accepted{...}

\submitjournal{ApJ}

\shorttitle{Accretion onto stars in AGN disks}
\shortauthors{Dittmann, Cantiello, and Jermyn}

\begin{document}
\title{Accretion onto Stars in the Disks of Active Galactic Nuclei}

\correspondingauthor{Alexander J. Dittmann}
\email{dittmann@astro.umd.edu}

\author[0000-0001-6157-6722]{Alexander J.~Dittmann}
\affil{Center for Computational Astrophysics, Flatiron Institute, 162 5th Avenue, New York, NY 10010, USA}
\affil{Department of Astronomy and Joint Space-Science Institute, University of Maryland, College Park, MD 20742-2421, USA}

\author[0000-0002-8171-8596]{Matteo Cantiello}
\affil{Center for Computational Astrophysics, Flatiron Institute, 162 5th Avenue, New York, NY 10010, USA}
\affil{Department of Astrophysical Sciences, Princeton University, Princeton, NJ 08544, USA}

\author[0000-0001-5048-9973]{Adam S. Jermyn}
\affil{Center for Computational Astrophysics, Flatiron Institute, 162 5th Avenue, New York, NY 10010, USA}

\begin{abstract}
Disks of gas accreting onto supermassive black holes are thought to power active galactic nuclei (AGN).
Stars may form in gravitationally unstable regions of these disks, or may be captured from nuclear star clusters.
Because of the dense gas environment, the evolution of such embedded stars can diverge dramatically from those in the interstellar medium. 
This work extends previous studies of stellar evolution in AGN disks by exploring a variety of ways that accretion onto stars in AGN disks may differ from Bondi accretion.
We find that tidal effects from the supermassive black hole significantly alter the evolution of stars in AGN disks, and that our results do not depend critically on assumptions about radiative feedback on the accretion stream. Thus, in addition to depending on the ambient density and sound speed, the fate of stars in AGN disks depends sensitively on the distance to and mass of the supermassive black hole. This affects where in the disk stellar explosions occur, where compact remnants form and potentially merge to produce gravitational waves, and where different types of chemical enrichment take place.   
\end{abstract}


\section{Introduction} \label{sec:intro}
The centers of most massive galaxies are thought to harbor supermassive black holes (SMBHs) \citep{2013ARA&A..51..511K}. These SMBHs can power quasars and active galactic nuclei (AGNs) through the release of gravitational energy as matter spirals into their deep potential wells \citep{{1969Natur.223..690L},{2008ARA&A..46..475H}}. 
Analyses of the redshift-dependent AGN luminosity function have inferred the efficiency, luminosity, and integrated duration of SMBH accretion \citep{{1982MNRAS.200..115S},{2002MNRAS.335..965Y}}, suggesting that SMBHs typically accrete for a total of tens to hundreds of millions of years in luminous quasars, although individual episodes may be as short as $\sim 10^5$ years \citep{{2015MNRAS.453L..46K},{2015MNRAS.451.2517S}}.

AGN accretion has historically been modeled as occurring through geometrically thin, optically thick accretion disks \citep{{1973A&A....24..337S},{1973blho.conf..343N}}. Such disk models are thermally unstable at small radii where radiation pressure dominates  \citep{1976MNRAS.175..613S,1976ApJ...204..187S}, although radiation-magnetohydrodynamics simulations indicate that they may be stabilized by magnetic pressure or convection-driven turbulence \citep{{2019ApJ...885..144J},{2019ApJ...880...67J},{2020ApJ...900...25J}}. Additionally, standard thin accretion disk models \citep{1973A&A....24..337S} become gravitationally unstable at large radii, and multiple methods of taking this into account have been suggested \citep{{2003MNRAS.341..501S},{2005ApJ...630..167T}}. 

When compared to microlensing observations, classical thin disk models under-predict accretion disk sizes at optical wavelengths by factors of $\sim4$ at IR, optical, and UV wavelengths \citep{{2007ApJ...661...19P},{2010ApJ...709..278D},{2010ApJ...712.1129M},{2010ApJ...718.1079B}}. Additionally, disk models are typically steady state, and cannot account for AGN variability \citep{{1983ApJ...271..564A},{2009ApJ...698..895K},{2010ApJ...721.1014M}}, although phenomenological models have attempted to account for such behavior \citep{{2011ApJ...727L..24D},{2020arXiv201107151L}}.

Other AGN observations are also incompatible with standard models. Broad emission lines can suddenly appear or disappear, accompanied by dramatic changes in luminosity \citep{{2015ApJ...800..144L},{2019ApJ...874....8M},{2019ApJ...883...31F}}. Quasars exhibiting extreme variability systematically accrete at lower rates than typical quasars, although such quasars appear to belong to the tail of a continuous distribution, rather than to a distinct population \citep{2018ApJ...854..160R}. 
Additionally, spectroscopic modeling of AGN broad line regions indicates that their metallicitiy is generally greater than solar, roughly independent of redshift, and an increasing function of SMBH mass \citep{{2006A&A...459...85N},{2018MNRAS.480..345X}}. The iron abundance in AGN disks may also be substantial, evidenced by the observation of the iron $K\alpha$ X-ray emission line \citep{{1995Natur.375..659T},{1997ApJ...488L..91N}}, although abundance inferences are subject to substantial uncertainties \citep{{2018ApJ...855....3T},{2018ASPC..515..282G}}.

Furthermore quasars do not exist in isolation: although AGNs typically outshine their host galaxies, observations suggest that nuclear star clusters are ubiquitous around SMBHs \citep{2020A&ARv..28....4N}. During active phases, these stars can be captured into the accretion disk even if initially on misaligned orbits \citep[e.g.][]{{1993ApJ...409..592A},{1995MNRAS.275..628R}}. Passages through the disk act to circularize eccentric stellar orbits \citep[e.g.][]{1991MNRAS.250..505S, 2020ApJ...889...94M}, and torques can further align and circularize stellar orbits \citep{{1995MNRAS.275..628R},{2004ApJ...602..388T}}. Stars may also form in the outer regions of AGN disks, where the disk is gravitationally unstable and the cooling timescale is short \citep{{1964ApJ...139.1217T},{2001ApJ...553..174G}}. 

Stars in AGN disks (AGN stars) and the compact objects left behind at the end of their evolution can enhance AGN metallicities \citep{1993ApJ...409..592A}, source LIGO/VIRGO events \citep[e.g.][]{{2017ApJ...835..165B},{2017MNRAS.464..946S},{2018ApJ...866...66M},{2020ApJ...898...25T}}, and contribute to the growth of high-redshift SMBHs \citep{2020MNRAS.493.3732D}. However, the evolution of AGN stars is much more exotic than the evolution of typical stars in the interstellar medium, due to the high ambient density and temperature in AGN disks. This alters the stellar boundary conditions and hence their structure and evolution.

Previous works have studied stellar evolution subject to AGN-like boundary conditions, ranging from stars subject to irradiation from an AGN \citep{1989MNRAS.238..427T} to those embedded within AGN disks and subject to Bondi-like spherically symmetric accretion and super-Eddington mass loss \citep{2020arXiv200903936C}. In this work we consider alternative models of radiation feedback on accretion, as well as deviations from spherical symmetry inherent to a disk environment, described in Section \ref{sec:analytic}. We outline our numerical methods in Section \ref{sec:numerical}, and present results in Section \ref{sec:results}. We find that the assumptions of radiative feedback have minor effects on stellar evolution in AGN disks, and that tidal effects are the most important of those considered in this work. We show how chemical enrichment varies as a function of generic disk parameters, and give general formulas for which accretion disk conditions lead AGN stars to end their lives as explosive transients, or live indefinitely with mass loss and accretion balancing one another.

\section{Analytic Considerations}\label{sec:analytic}

When the specific angular momentum of accreting material is low, and the radiation from the accreting star is sufficiently sub-Eddington, accretion may be modelled as a spherically symmetric process occurring at the Bondi rate
\begin{equation}\label{mBondi}
\dot{M}_B = \eta\pi R^2_B\rho c_s,
\end{equation}
where $\eta$ is an efficiency factor ($\eta\lesssim 1$), $\rho$ and $c_s$ are the density and sound speed of ambient gas,
\begin{equation}\label{rBondi}
R_B = \frac{2GM_*}{c_s^2}
\end{equation}
is the Bondi radius, $M_*$ is the mass of the star and $G$ is the gravitational constant \citep{1952MNRAS.112..195B}.
 These equations assume that the relative velocity between the star and ambient medium is much less than the sound speed. Retrograde stellar orbits have a relative velocity to the gas disk $\Delta v \gg c_s$, so Equations (\ref{mBondi}) and (\ref{rBondi}) are not applicable in this scenario \citep[see, e.g.][]{1939PCPS...35..592H}. For prograde orbits, the relative velocity comes from Keplerian shear and pressure gradients, and becomes larger than the sound speed when the Bondi radius is larger than the disk scale height, in which case the assumption of symmetry breaks down and Equation (\ref{mBondi}) is no longer appropriate.

In addition to the Bondi radius, two other length-scales are important for stars in AGN disks.
First, the scale height of the disk is given by
\begin{align}
H \equiv \sqrt{2}\,\frac{c_s}{\Omega},
\end{align}
where
\begin{align}
    \Omega = \sqrt{\frac{G M_\bullet}{r_\bullet^3}}
\end{align}
is the Keplerian angular velocity of the AGN disk, $M_\bullet$ is the mass of the SMBH, and $r_\bullet$ is the distance from the star to the SMBH.
Second, the Hill radius \citep{hill1878researches} is
\begin{equation}
R_H = r_\bullet \left(\frac{M_*}{3M_\bullet} \right)^{1/3}=\left(\frac{GM_*}{3\Omega^2} \right)^{1/3}.
\end{equation}
This is the radius of a sphere within which the gravity of the star dominates that of the SMBH.
Note that with the above definitions $R_H/R_B \propto H^2/R_H^2$ and $H/R_H\propto hq^{-1/3}$, where $q\equiv M_*/M_\bullet$ is the mass ratio and $h\equiv H/r_\bullet$. 

Hereinafter, we make the approximation that the gas accreting onto AGN stars has a uniform sound speed, and either a uniform density ($\rho=\rho_0$) or one that varies only as a function of height from the midplane ($\rho=\rho_0 f(z)$), where $\rho_0$ is the midplane gas density. Although this approximation should break down in most cases, especially as the length scale for accretion becomes large, it enables us to keep our investigation largely independent of the precise and highly uncertain structure of AGN disks. 

\subsection{Radiative feedback}\label{sec:rad}
As stars become more massive, their luminosity ($L_*$) increases rapidly, $L_*\propto M_*^{\sim3}$~\citep{1992isa3.book.....B}. As this radiation impinges on ambient gas, it causes a specific force on gas a distance $r$ away of $f_g = F\kappa/c$, where $\kappa$ is the opacity of the gas, $c$ is the speed of light, and $F$ is the radiative flux, given for a spherically symmetric radiation field by $L_*/4\pi r^2$. The net acceleration of a gas parcel due to radiation and gravity is thus
\begin{equation}\label{fBalance}
f=\frac{1}{r^2}\left(\frac{L_*\kappa}{4\pi} - GM_*\right) = -\frac{GM_*}{r^2}\left(1 - \frac{L_*\kappa}{4\pi GM_*c}\right).
\end{equation}
The luminosity where the net acceleration becomes zero is the Eddington luminosity
\begin{align}
L_{\rm Edd}=4\pi GM_*c/\kappa.
\end{align}
Moreover, from Equation (\ref{fBalance}), it can be seen that the \textit{effective} mass of the star as experienced by ambient gas is reduced by a factor of $(1-L_*/L_{\rm Edd})$. As $R_B\propto M_*$, the \textit{effective} Bondi radius is reduced by the same factor. Similarly, the accretion rate onto a star is modified according to
\begin{equation}\label{radiation2}
\dot{M}=\dot{M}_B\left(1-\frac{L_*}{L_{\rm Edd}}\right)^2.
\end{equation}

This picture is entirely one dimensional, which is a drastic simplification of reality.
For example, rotating stars are more luminous at their poles than their equators \citep{{1924MNRAS..84..665V},{1967ZA.....65...89L}}, and density perturbations can lead to channels of accretion through the Rayleigh-Taylor instability \citep{{2013MNRAS.434.2329K},{2014ApJ...796..107D}}. \citet{2020arXiv200903936C} attempted to account for deviations from spherical symmetry by using, instead of Equation (\ref{radiation2}), the following phenomenological prescription:
\begin{equation}\label{radiation1}
\dot{M} = \dot{M}_B \left(1 - \tanh{|L_*/L_{\rm Edd}|}\right).
\end{equation}
This prescription is useful because it allows super-Eddington accretion, which can happen in geometries where radiation is able to escape in one direction while accretion primarily happens along another, among other scenarios. We present results computed using each prescription (Equations~\eqref{radiation2} and~\eqref{radiation1}) and find that they lead to quantitative, but not qualitative, changes in the overall picture.

\begin{figure}[h!]
\centering
\includegraphics[width=.992\columnwidth]{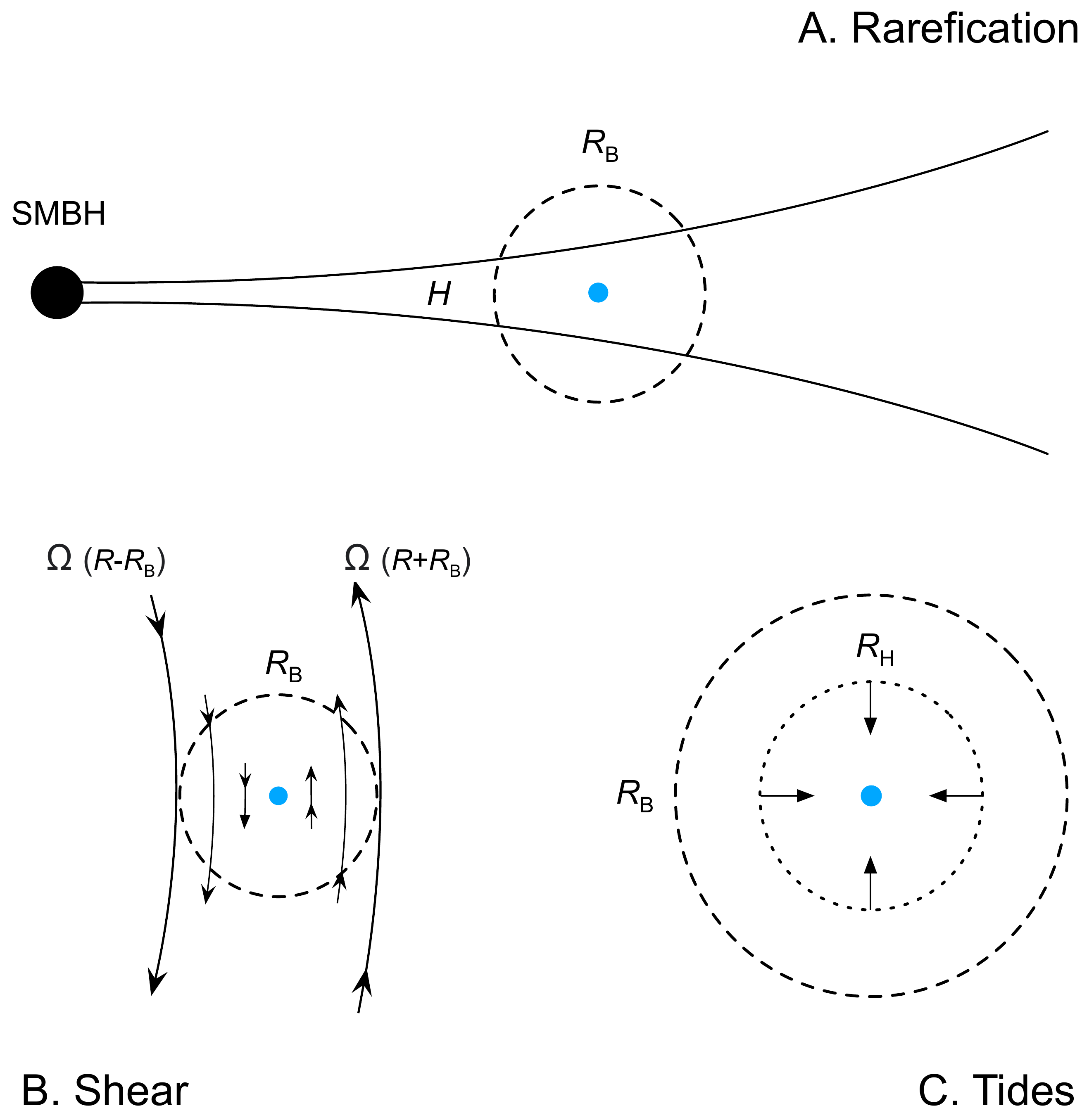}
\caption{Accretion onto an AGN star (blue circle) orbiting a supermassive black hole can be limited by geometric and tidal effects. Geometric effects include rarefication (A), which is important when the Bondi radius $R_{B}$ becomes comparable to or larger than the disk's scale height $H$, and shear (B), which is due to the Keplerian rotation of the disk. Tidal effects (C) are caused by the gravity of the SMBH, which becomes important when the Hill radius is smaller than the Bondi radius.}
\label{sketch}
\end{figure}

\subsection{Vertical stratification}
It is common to assume that accretion disks are vertically isothermal.
In such disks, vertical hydrostatic equilibrium leads to
$\rho(z)=\rho_0\exp{\left[-(z/H)^2\right]}$ when $r\gg z$.
For a star in the midplane of the disk this means that the average density of gas at the Bondi radius decreases with increasing Bondi radius, and so generally decreases as the star becomes more massive.
Deviations from an isothermal structure or a more realistic treatment of radiation transport and opacities could reveal density inversions or more shallow density gradients \citep[e.g.][]{{1982A&A...106...34M},{1994ApJ...421..668M},{2000ApJ...533..710H},{2020ApJ...900...25J}}. Because these complications all reduce the effect of stratification, assuming a vertically isothermal disk allows us to gauge the maximum impact that vertical density variations could have on the evolution of AGN stars. 

To account for how changes in density vertically can decrease the accretion rate, we average the density in Equation (\ref{mBondi}) over a sphere with radius $R_B$ centered on the star. We find that in this case, with $z=R_B\cos{\theta}$ in a polar coordinate system centered on the star, 
\begin{equation}
\begin{split}
\frac{\langle \rho \rangle}{\rho_0} = \frac{1}{2}\int_0^\pi\sin{(\theta)} \exp\left[-\left(\frac{R_B}{H}\right)\cos^2{(\theta)}\right]d\theta \\= \frac{\sqrt{\pi}}{2}\frac{H}{R_B}{\rm erf}\left(\frac{R_B}{H}\right),
\end{split}
\end{equation}
where $\langle ... \rangle$ denotes an average over the Bondi sphere, $\rm erf$ is the error function and $\rho_0$ is the density of the midplane of the disk.
If we use $\langle \rho \rangle$ to compute the accretion rate we then find
\begin{equation}\label{rarify}
\dot{M}= \dot{M}_B\frac{\sqrt{\pi}}{2}\frac{H}{R_B}{\rm erf}\left(\frac{R_B}{H}\right).
\end{equation}

\subsection{Shear} \label{sec:shear}
Shear in the accretion disk imbues accreting gas with net angular momentum in the frame comoving with the star, and can therefore limit accretion onto an embedded star. 
To leading order, at a distance $\Delta s$ away from the star the linear velocity ($v=r_{\bullet}\Omega)$ of the disk is different from that of the star by an amount
\begin{align}
    \Delta v = v(r_\bullet) - v(r_\bullet + \Delta s)\approx \frac{1}{2}v(r_\bullet)\frac{\Delta s}{r_\bullet} = \frac{1}{2}\Omega(r_\bullet)\Delta s,
\end{align}
where we have assumed a Keplerian rotation profile for simplicity.
In a spherical coordinate system centered on the star, $\Delta s = R_B \cos{(\theta)}\sin{(\phi)}$ at a distance $R_B$ from the star. Averaging the specific angular momentum relative to the star, $l = \Delta s \Delta v$, over the Bondi sphere, we find that the average specific angular momentum is 
\begin{equation}
\begin{split}
\langle l \rangle  = R_B^2\frac{\Omega(r_\bullet)}{8\pi}\int_0^{2\pi}\sin^2{(\phi)}d\phi\int_0^\pi\cos^2{(\theta)}\sin{(\theta)}d\theta\\
=R_B^2\frac{\Omega(r_\bullet)}{12}.
\end{split}
\end{equation}

Following \citet{2005ApJ...618..757K}, we estimate the impact of angular momentum on the accretion. Let
\begin{align}
\omega_*=\frac{\langle l \rangle}{c_s R_b},
\end{align}
where $\omega_*$ is a measure of the vorticity of the flow.
Then this angular momentum reduces the accretion rate by a factor of
\begin{equation}\label{mShear}
\frac{\dot{M}}{\dot{M}_B} = \min\{1, f_s(\omega_*)\},
\end{equation}
 where 
\begin{equation} \label{shearacc}
f_s(\omega_*) = \frac{2}{\pi \omega_*}\sinh^{-1}\left[(2\omega_*)^{1/3}\right],
\end{equation}
Equation (A7) of \citet{2005ApJ...618..757K}. Note that in this case, $\omega_*=R_B\Omega/12c_s=R_B/6\sqrt{2}H$. Since this factor does not become significant until the Bondi radius is more than eight times the disk scale height, shear can be expected to affect accretion less than other factors such as rarefication.

\subsection{Tidal effects}
When the Bondi radius is much smaller than the disk scale height and Hill radius, disk geometry and the gravity of the SMBH have a negligible effect on accretion onto AGN stars.
However, because $R_B/R_H \propto M_\star^{2/3}$, $R_B$ naturally becomes larger than $R_H$ as stars grow more massive. The reason this alters the accretion rate onto stars can be understood by considering the case where
$R_B \ga R_H$. In this case, the motion of gas at $R_B > R > R_H$ is controlled by the gravity of the SMBH with only minor influences from the star.
To incorporate this effect we replace the Bondi radius with the smaller out of the Hill and Bondi radii when calculating accretion rates 
\begin{equation}
\label{tidalHigh}
\dot{M} = \dot{M}_B \min\left\{1, \left(\frac{R_H}{R_B}\right)^2 \right\}, 
\end{equation}
along the lines of \citet{2020MNRAS.498.2054R}.

In disks with sufficiently low viscosity, accretion can be further reduced by the so-called `tidal barrier', and by the conservation of potential vorticity along flow lines \citep{2007ApJ...660..791D}. \citet{2021ApJ...906...52L} carried out a suite of 2D and 3D viscous hydrodynamics simulations using an $\alpha-$viscosity prescription. For $\alpha \ga 10^{-2},$ the results were better described by the scaling of \citet{2020MNRAS.498.2054R} (equation~\eqref{tidalHigh} above), while for lower $\alpha$ the scalings of \citet{2007ApJ...660..791D} were a better fit. To the extent that an $\alpha$-viscosity accurately describes angular momentum transport in AGN disks, simulations and comparisons with observations suggest effective values of $\alpha$ in AGN disks larger than $10^{-2}$ \citep{{2016ApJ...826...40H},{2007MNRAS.376.1740K}}.
Hence we expect that (\ref{tidalHigh}) is appropriate for stars in AGN disks. 

Recall also that the strength of tidal effects scales strongly with both the star's radius ($R_*$) and the distance between the star from the SMBH. For example, as discussed in \citet{doi:10.1063/1.45957}, tidal dissipation in stars captured into eccentric orbits scales as $\dot{E}_{\rm tidal}\propto M_\bullet^{5/2}R_*^{5}r_\bullet^{-15/2}.$ For the conditions considered in this work, tidal heating is typically small, $\dot{E}_{\rm tidal}<L_\odot$. Similarly, tidal dissipation does not significantly affect the orbital evolution of the stars considered here. However, stars could grow to have larger masses and radii in different gas conditions, and tidal dissipation becomes much stronger close to the SMBH. These caveats should be kept in mind, especially when considering stellar evolution very near the SMBH.

\subsection{Gap opening}
If stars become sufficiently massive, they can open a gap in the disk due to strong nonlinear disk-star interactions \citep{1997Icar..126..261W, 2014ApJ...792L..10D, 2018ApJ...861..140K}. Contrary to the classical picture of gap opening, where gas cannot flow into the gap and the object is locked in step with the viscous evolution of the disk \citep{1997Icar..126..261W}, several numerical studies have shown that gas can flow into the gap, albeit at a much lower surface density, and that objects within the gap can migrate faster or slower than the viscous speed of the disk \citep{2014ApJ...792L..10D, 2015A&A...574A..52D}. The decrease in disk surface density can lead to a commensurate decrease in the accretion rate onto the star. The degree to which the surface density of gas in the gap is depleted is described fairly well by 
\begin{equation} \label{depleted}
\frac{\Sigma_{\rm min}}{\Sigma}=\frac{1}{1+0.04K},
\end{equation}
where $K=q^2h^{-5}\alpha^{-1}$ \citep{{2015MNRAS.448..994K},{2015ApJ...807L..11D}}. Equation (\ref{depleted}) follows from the assumptions that the gap depth is determined by the balance between torques from the disk and the viscous angular momentum flux, along with the assumption that the object primarily interacts with gas at the bottom of the gap, which was found to be consistent with the results of various hydrodynamical simulations \citep{2017PASJ...69...97K}.

For $K\gtrsim10$, the surface density of the disk can be significantly depleted, further decreasing the accretion rate onto massive AGN stars. Because we expect modest values of $\alpha$ ($>10^{-2}$)  to be applicable to AGN disks, as opposed to $\alpha < 10^{-4}$ as may be applicable in planetary contexts \citep{2013ApJ...769...41D}, even for values of $h \sim 10^{-2}$, $q$ would need to be greater than $10^{-3}$ for the disk surface density to deplete significantly. Thus, even for low-mass SMBHs with $M_\bullet\sim10^6~\Msun$, stars would need to reach $M_* \gtrsim 10^3~\Msun$ before beginning to meaningfully influence the disk surface density. 
We find that AGN stars typically reach masses at most $\sim 10^3~\Msun$, so we do not consider the effects of gap formation on accretion in this work. However, gap opening could significantly alter the evolution of stars orbiting lower-mass black holes.

\section{Numerical Methods}\label{sec:numerical}
We model the evolution of AGN stars using revision 15140 of the Modules for Experiments in Stellar Astrophysics \citep[MESA][]{{2011ApJS..192....3P},{2013ApJS..208....4P},{2015ApJS..220...15P},{2018ApJS..234...34P},{2019ApJS..243...10P}} software instrument. 
We implement mass loss and modified boundary conditions following Cantiello et al. (2020). These boundary conditions include a treatment of stellar irradiation as well as e.g. the ram pressure of the accretion stream.
We consider accretion of a gas mixture with mass fractions $X=0.72,$ $Y=0.28$, and $Z=0$. Although this is not realistic for gas in most AGN disks, it highlights metal production by AGN stars and allows us to successfully evolve stars without incurring numerical issues. 
The dependence of AGN star evolution on accreted composition is left for the future.

In addition to the accretion prescription used in  \citet{2020arXiv200903936C}, we implement in MESA the various processes described in Section \ref{sec:analytic}. One of them, Equation (\ref{rarify}), requires special treatment to ensure reproducibility and make use of the auto-differentiation module within MESA. We approximate the error function by a B\"urmann series \citep{{1963cma..book.....W},{Schopf2014}}
\begin{equation}
\begin{split}
{\rm erf}(x)\approx \frac{2}{\sqrt{\pi}}\frac{x}{|x|}\sqrt{1-e^{-x^2}}\left(\frac{\sqrt{\pi}}{2}+\frac{31}{200}e^{-x^2}-\frac{341}{8000}e^{-2x^2}  \right),
\end{split}
\end{equation}
which has a maximum relative error $\lesssim 0.37\%$ and produces, bit-for-bit, the same result on all computational platforms owing to the use of the CR-LIBM library~\citet{CR-LIBM}.

As shown by \citet{2020arXiv200903936C}, many AGN stars evolve with $L_*\sim\ledd$, such that super-Eddington continuum-driven winds are likely to dominate their mass loss. Therefore, we do not include line-driven mass loss \citep[but see][for a review of the different ways massive stars can lose their mass]{2014ARA&A..52..487S}. Similar to other works \citep[e.g.][]{2011ApJS..192....3P}, we assume a super-Eddington outflow at the escape velocity $v_{\rm esc} = (2GM_*/R_*)^{1/2}$, and following~\citet{2020arXiv200903936C} we compute an associated mass-loss rate
\begin{equation}\label{loss}
\dot{M}_{\rm loss}=-\frac{L_*}{v_{\rm esc}^2}\left[1 + \tanh{\left(\frac{L_*-\ledd}{0.1\ledd} \right)} \right],
\end{equation}
where the hyperbolic tangent acts to smooth the onset of mass loss and help our calculations converge.

For models approaching the Eddington luminosity we also assume an enhancement of compositional mixing in radiative regions. This is justified by the fact that the threshold for vertical instability decreases as 
stars become radiation dominated and the adiabatic index $\Gamma_1$ approaches $4/3$. A number of processes can trigger instabilities in the radiative envelope, including rotation~\citep{1929MNRAS..90...54E} and locally super-Eddington luminosities~\citep{2018arXiv180910187J}. Details of the mixing implementation are discussed in \citet{2020arXiv200903936C}.

We initialized our runs with a zero-age main sequence solar model.
We then relaxed the boundary conditions and accretion rate over approximately $10^7~\mathrm{yr}$ from solar-like to those described by~\citet{2020arXiv200903936C}.
This relaxation happens sequentially, with the boundary conditions fully relaxed before the accretion rate relaxation begins.
The total relaxation time is kept short enough that it does not amount to a significant fraction of the total evolutionary time of our initial $1~\Msun$ models.

\section{Results}\label{sec:results}
\subsection{Archetypal Models}

It is useful to examine a few archetypal evolutionary tracks before studying whole populations.

\begin{figure}[h!]
\centering
\includegraphics[width=.992\columnwidth]{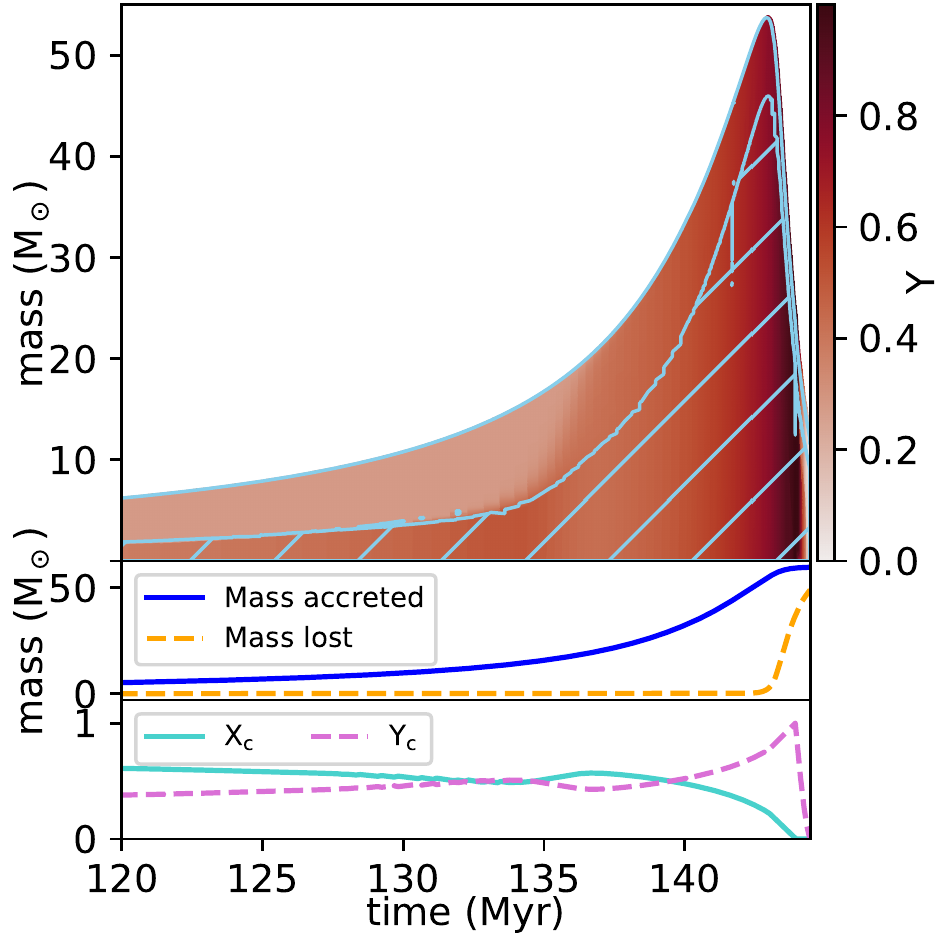}
\caption{The evolution of a typical model in the intermediate accretion regime is shown. The ambient gas density is $2 \times 10^{-18}~\rm g~\rm{cm}^{-3}$ and the sound speed is $10~\rm{km}~s^{-1}$. The top panel shows the helium mass fraction in the star as a function of time and stellar mass coordinate. Convective regions are marked by light blue hatching. The middle panel displays the mass accreted and mass lost over time. The lower panel shows the core hydrogen and helium mass fractions throughout the star's life.}
\label{intermediate}
\end{figure}

\begin{figure}[h!]
\centering
\includegraphics[width=.992\columnwidth]{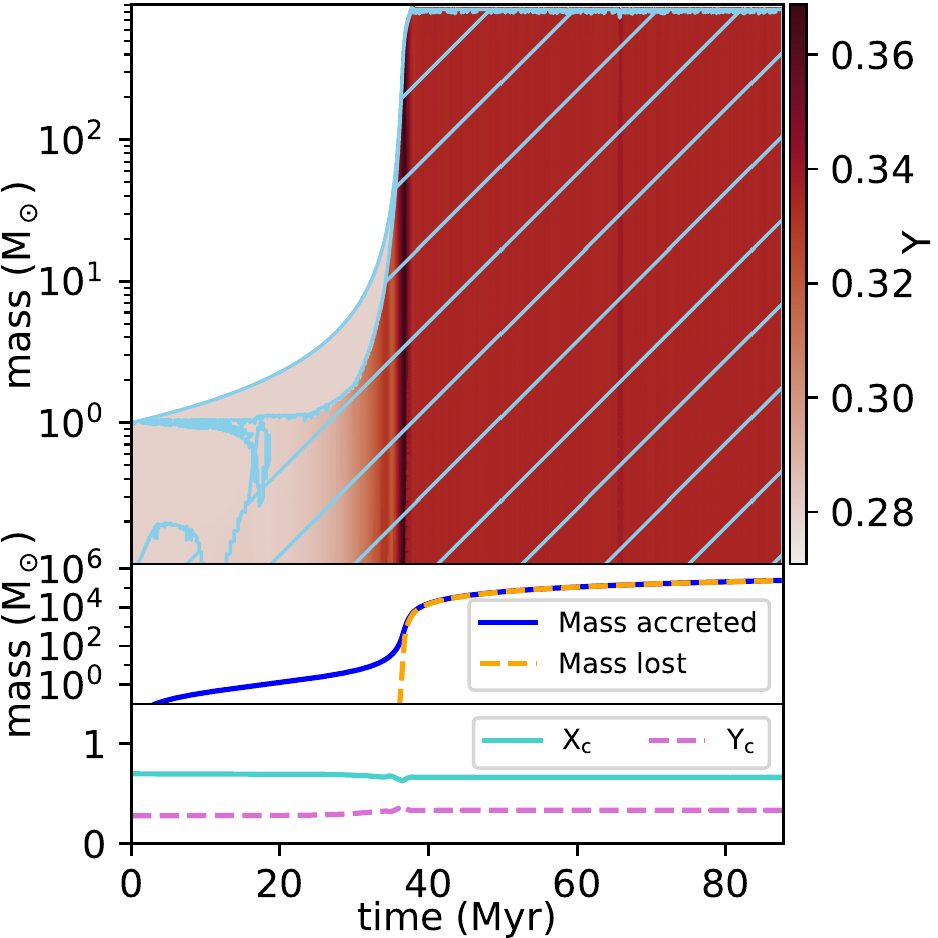}
\caption{The evolution of a typical model in the runaway accretion regime, leading to an ``immortal'' AGN star. The ambient gas density is $8 \times 10^{-18}~\rm g~\rm{cm}^{-3}$ and the sound speed is $10~\rm{km}~s^{-1}$. The top panel shows the helium mass fraction in the star as a function of time and stellar mass coordinate. Convective regions are marked by light blue dashed lines. The middle panel displays the mass accreted and mass lost over time. The lower panel shows the core hydrogen and helium mass fractions throughout the star's life. }
\label{immortal}
\end{figure}

In cases where the accretion rate is very small\footnote{Accretion timescale much longer than the stellar nuclear burning timescale.}, stellar evolution is largely unaffected and stars reach the end of their lives with a similar mass to that at which they began. However, stars subject to higher accretion rates can have significantly different evolutions.
This difference can be seen in Figures~\ref{intermediate} and~\ref{immortal}, which show the evolution of a $1\,\Msun$ AGN star embedded in gas with a sound speed of $10~\rm{km}~s^{-1}$ and ambient gas densities of $2 \times 10^{-18}~\rm g~\rm{cm}^{-3}$ and $8 \times 10^{-18}~\rm g~\rm{cm}^{-3}$ respectively. 

The stars begin with a steep composition gradient, but become chemically homogeneous as they increase in mass, accreting fresh hydrogen and helium. Throughout most of its life,  the $\rho=2 \times 10^{-18}~\rm g~\rm{cm}^{-3}$ model has a convective core and radiative envelope. From about $\sim 134 - 137$ Myr, the hydrogen mass fraction in the core increases due to mixing of accreted material. After this, accretion is effectively halted by radiative feedback and the star loses its source of fresh hydrogen.
The helium mass fraction then increases throughout the star until it is eventually depleted through fusion into heavier elements. This general trend is followed by stars in the `intermediate' accretion regime, where the nuclear burning timescale is shorter than, but comparable to, the accretion timescale \citep{2020arXiv200903936C}.

Unlike the $\rho=2\times10^{-18}~\rm g~\rm cm^{-3}$ case, in the  $\rho=8\times10^{-18}~\rm g~\rm cm^{-3}$ model mass loss and accretion come into an approximate balance once the star nears the Eddington limit. As a consequence the mass of the model reaches an approximate steady state. The star is almost fully convective, so accreted material is rapidly mixed throughout. For the same reason, mass lost from the star is chemically enriched by the ashes of nuclear fusion occurring in the stellar core. Examining the mass loss and accretion budget, it is evident that this single star that began at $1~\Msun$ is able to process well above $10^5~\Msun$ worth of gas in $\sim10$s of Myr (Figure~\ref{immortal}, middle panel). Stars with higher initial masses or in higher-density (or lower-sound speed) environments are able to reach this evolutionary stage much more quickly, since $M/\dot{M}_B \propto M^{-1}\rho^{-1}$.
Because these models can persist at high masses indefinitely in the appropriate conditions, we refer to these models as `immortal' for simplicity. Such models are the result of accretion in the `runaway' regime discussed in \citet{2020arXiv200903936C}, where the accretion timescale is less than or equal to the burning timescale.

\begin{figure}[h!]
\centering
\includegraphics[width=.992\columnwidth]{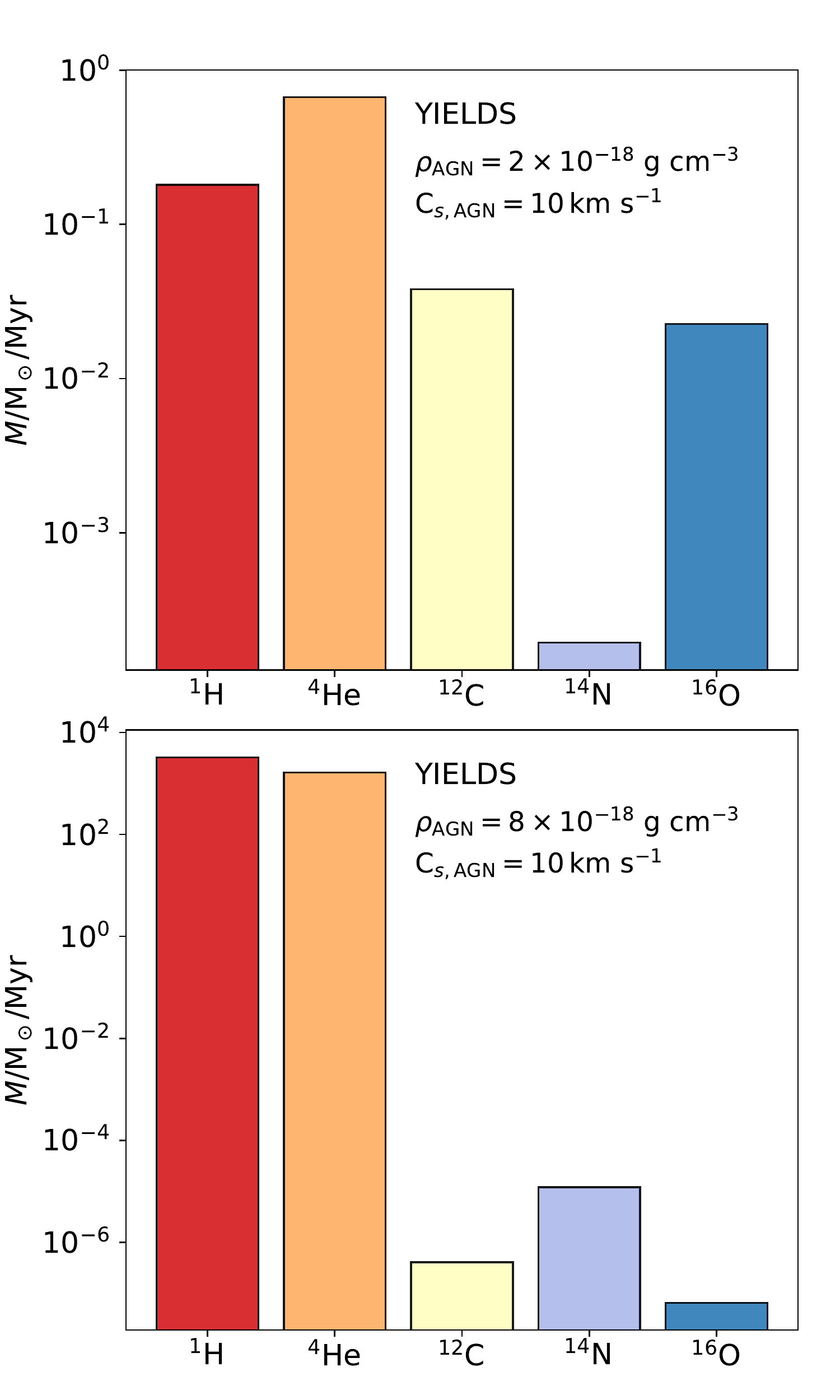}
\caption{Chemical yields for archetypal `intermediate' (figure \ref{intermediate}, upper) and `immortal' (figure \ref{immortal}, lower) models. For both cases, we have excluded the first $4\times10^7$ years of the star's lives. Because the immortal model could fuse indefinitely, we have presented yields in terms of the mass lost per million years. Note that the yields from immortal stars contain very little metal content because these stars primarily fuse hydrogen, although their metal content is significantly enriched in nitrogen. As mortal stars lose mass and reach later stages of evolution they eventually produce significant amounts of carbon and oxygen as well.}
\label{exampleYields}
\end{figure}

The chemical yields of mass lost from these models are shown in Figure \ref{exampleYields}. Although the immortal model loses mass at a rate $\sim 10^4$ times larger than the intermediate model, this material has a much lower metal content and so the overall yield of metals is lower. Thus, mass loss from immortal stars matters less for enriching the disk with metals than that lost from 
intermediate stars, although the large mass loss rates could play an important role in determining the overall structure of the AGN disk. Additionally, stars may migrate through the disk, and the disk itself will dissipate over time. Thus, eventually `immortal' stars will find themselves in lower-density environments and begin to evolve similarly to those in the intermediate accretion regime, reaching later stages of burning and losing most of their nuclear-processed material \citep[e.g. Figure 10 and 11 in][]{2020arXiv200903936C}.

\subsection{Model Grids}

\begin{figure*}[h]
\centering
\includegraphics[width=.992\textwidth]{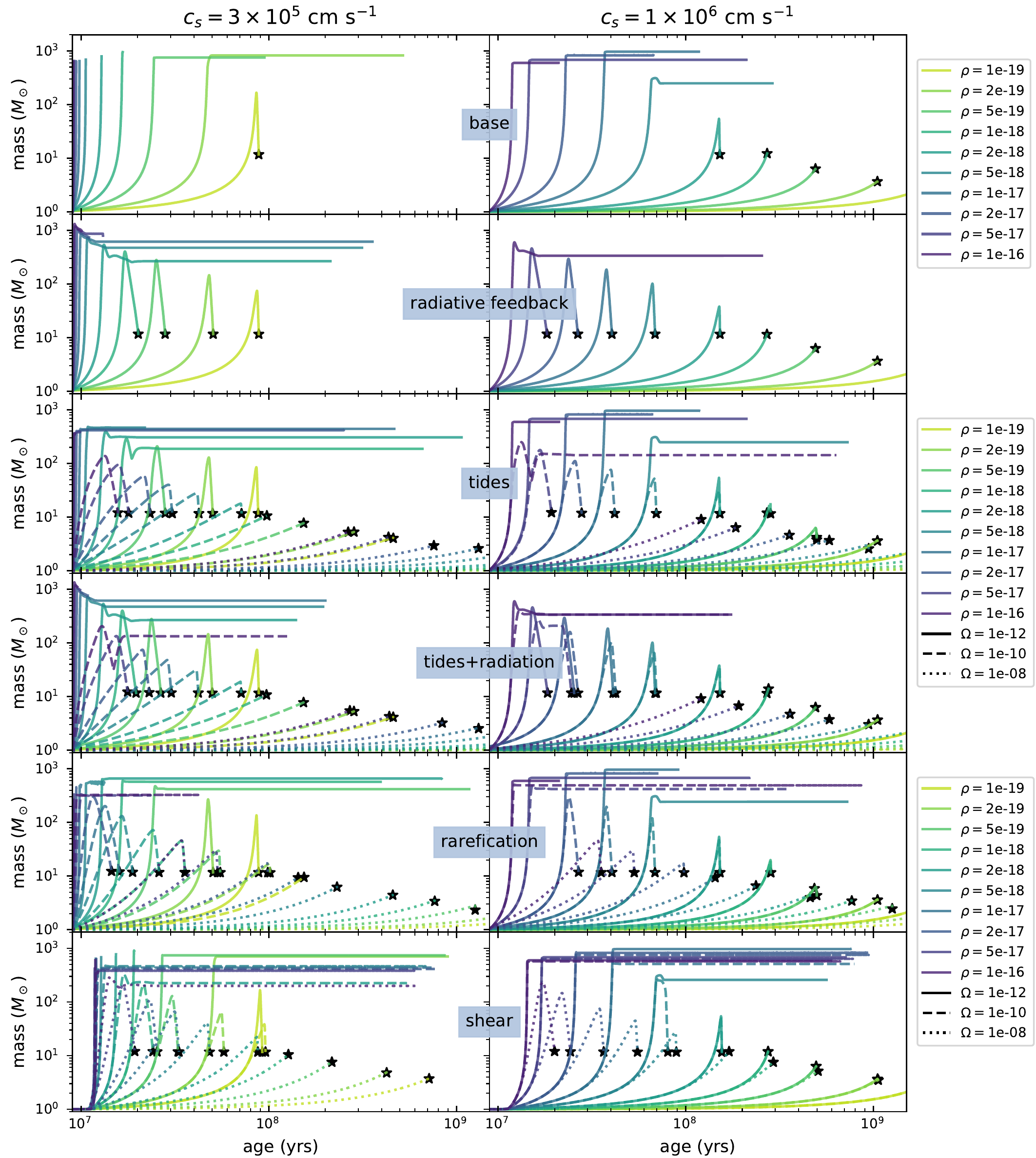}
\caption{Evolutionary tracks of stellar mass over time for various accretion prescriptions. On the x-axes, the reported age includes the $\sim10^7$ years over which the boundary conditions for each model were relaxed. We begin plotting results at the end of the relaxation process for each run. Each column corresponds to a different ambient sound speed, as indicated by the column headers. Each row presents results for a different accretion prescription, as indicated by the text boxes in each row. Line colors correspond to different ambient gas densities (given in $\rm{g~cm^{-3}}$), and line styles correspond to different AGN disk angular frequencies (given in $\rm{s^{-1}}$) at the star's location. Stars indicate the final masses of stars at the end of their lives.  Recall that for $M_\bullet =10^8~\Msun$, $\Omega = \{10^{-12},10^{-10},10^{-8}\}~\rm s^{-1}$ corresponds to distances from the SMBH $r_\bullet \approx \{7.67, 0.356, 0.0265\}~$pc.
}\label{manypresc}
\end{figure*}

With these general types of evolution in mind, we aim to understand how each modification to accretion discussed in Section \ref{sec:analytic} affects the evolution of AGN stars. We present in Figure~\ref{manypresc} a series of $M_*(t)$ diagrams for each of the accretion prescriptions discussed in Section \ref{sec:analytic}.
The prescriptions used by~\citet{2020arXiv200903936C} are taken as a baseline and shown in the first row.
Each subsequent row demonstrates exactly one modification from baseline, with the exception of the fourth row which includes two modifications. 
In each panel we show a variety of $\Omega$, $\rho$ (the angular velocity and density of the gas at the stellar location in the AGN disk).
Colors show density, with lighter and yellower colors indicating stars in lower-density environments and darker and bluer colors indicating those in higher-density environments. The models in the left column are embedded in gas with a sound speed of $3~\rm km~s^{-1}$, while the models in the right column are embedded in gas with a sound speed of $10~\rm km~s^{-1}$. Line styles indicate the angular velocity $\Omega$ about the SMBH, and where necessary $M_\bullet=10^8~\Msun$ is used to calculate $r_\bullet$. Note that some models do not depend on $\Omega$. This sparse grid of models elucidates the impact of the different modifications to accretion onto AGN stars.

We begin by considering the first and second rows, which show models computed using our baseline prescription and those computed using stronger radiative feedback (Equation~\eqref{radiation2}).
The more severe reduction in accretion rate imposed by Equation \ref{radiation2} extends the range of densities leading to intermediate stellar evolution, as opposed to runaway, by about an order of magnitude.

Of the effects we consider, tidal forces from the SMBH are generally the most significant. Specifically, tidal effects slow accretion onto AGN stars enough to shift many from the runaway regime to the intermediate regime, and to shift many from the intermediate regime into the regime of fairly standard stellar evolution.
Because of its independence from disk structure, support from hydrodynamical simulations \citep{2021ApJ...906...52L}, and having the greatest efficacy of the standalone modifications to the models in \citet{2020arXiv200903936C}, the tidal accretion prescription is promising for general use. One feature of tidally-limited accretion is that the accretion rate becomes proportional to the sound speed, $\dot{M}\propto R_H^2c_s\rho$, unlike Bondi accretion where $\dot{M}_B\propto c_s^{-3}$. Because of this, when tidal effects dominate, stars in gas with higher sound speeds tend to have larger final masses or maximum masses, as well as immortal-to-intermediate transitions at lower densities.

Because tidal effects depend on the ratio $R_H/R_B$, their impact may become less significant if $R_B$ is reduced by radiation from the star. To test the extent to which this occurs, we carried out another suite of simulations using both modifications. For small $\Omega$, tidal effects are minimal and stars evolve essentially identically to the models without tidal effects. However, at higher $\Omega$, 
tidal effects dominate and evolution is largely the same as when only considering tidal effects.
Thus we find that the evolution of AGN stars only depends significantly on raditative feedback assumptions for stars that are far from the SMBH, e.g. $\sim 0.36$ pc for a $10^8~\Msun$ black hole. 

Vertical stratification in the disk can also decrease the accretion rate onto AGN stars enough to cause qualitative deviations in their evolution. However, its effects are less significant than tidal effects because rarefication only operates in one dimension in our models. This can be understood by considering the growth of the Bondi radius as the star accretes: although gas directly above and below the star decreases in density significantly, gas near the midplane is relatively unchanged. Still, smaller disk scale heights can lead to greatly reduced accretion. We find that rarefication is also more significant for models accreting cooler gas. This can be understood by recalling that $H\propto c_s \Omega^{-1}$, so while $\Omega=10^{-10}~\rm{s^{-1}}$ and $\Omega=10^{-12}~\rm{s^{-1}}$ models evolve very similarly at $c_s=10^{6}~\rm{cm~s^{-1}}$, model tracks diverge significantly at $c_s=3\times10^{5}~\rm{cm~s^{-1}}$. 

We expect, however, that this investigation likely overestimates the effects of disk rarefication on the evolution of AGN stars, since AGN disks are not perfectly vertically isothermal, which would lead to a less steep decline in density vertically. Similarly, radiation-MHD simulations of AGN accretion disks tend to develop vertical profiles that decline less rapidly from the midplane than would a vertically isothermal disk \citep[e.g.][]{2019ApJ...885..144J}. Additionally, when taking into account realistic opacities, some disk models \citep{2000ApJ...533..710H} and simulations \citep{2020ApJ...900...25J} have shown that in the presence of opacity bumps, density inversions can occur in the disk, leading to an increase in density away from the midplane. Thus, the effects of rarefication here are likely to be unrealistically strong. 

We find shear to have a fairly negligible effect on the evolution of AGN stars. Because the effects of shear depend on the disk scale height, they lead to minor changes in the evolution of stars at $c_s=10^{6}~\rm{cm~s^{-1}}$ only fairly close to the SMBH, for $\Omega\gtrsim10^{-8}~\rm s^{-1}$. In cooler disks, with commensurately small scale heights, shear can be significant over a somewhat larger range in $\Omega$. As discussed in Section \ref{sec:shear}, shear becomes significant at much larger $R_B/H$ than for other effects such as tides or vertical stratification, so its lower efficacy at slowing accretion is natural.

\begin{figure}
\centering
\includegraphics[width=.992\columnwidth]{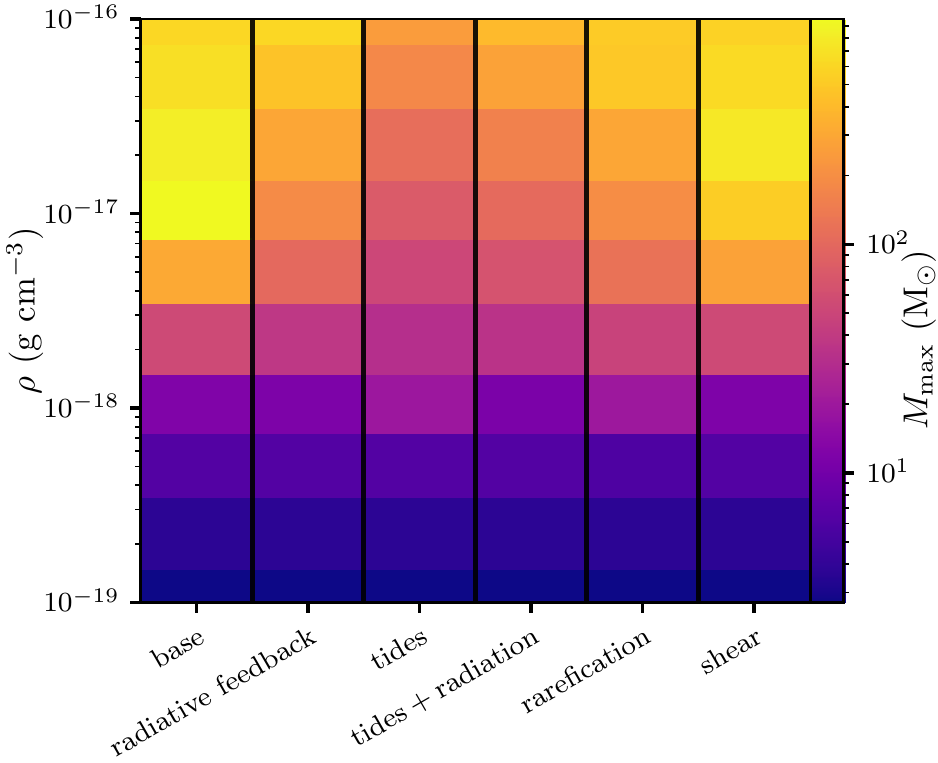}
\caption{The maximum mass reached for a subset of the models in Figure \ref{manypresc} with $c_s=10^6~\rm{cm~s^{-1}}$ and, where applicable, $\Omega = 10^{-10}~\rm{s^{-1}}$. Gray squares indicate models that failed during boundary condition relaxation. The y-axis indicates the disk density used for each model. Vertical black lines separate the results based on the accretion prescription used, indicated on the x-axis.}
\label{mass2dpresc}
\end{figure}

A summary of the maximum masses reached in the  $c_s=10^6~\rm{cm~s^{-1}},~\Omega=10^{-10}~\rm{s^{-1}}$ runs is presented in Figure \ref{mass2dpresc}. 
At this $\Omega$, models accounting for shear have almost no deviation from baseline calculations. Similarly, runs accounting for the rarefication of the disk reach maximum masses in between the masses reached by models using the base assumptions and those accounting for tidal effects. At this $\Omega$, the models accounting for reductions in the $R_B$ and $R_H$ due to radiation (as outlined in section \ref{sec:rad}) as well as tidal effects reach maximum masses that are lower than those of models that reduced $R_B$ but did not include tidal effects, and maximum masses that are higher than when accounting for tidal effects without reducing $R_H$ and $R_B$.

\subsection{Tide-mediated stellar evolution}
Having identified tides as the most significant modification to Bondi accretion, of those considered here, we now turn to study their effects on the evolution of AGN stars.
Recall that the Hill radius of a star can be written in terms of its mass and the angular velocity of its Keplerian orbit.
Similarly, the tide-limited accretion rate depends on, apart from stellar mass, $\rho c_s \Omega^{-4/3}$
Thus, to the extent that tides and accretion govern the evolution of AGN stars, we can study their evolution across a variety of disk characteristics and SMBH masses using only two parameters. 
For this reason we perform the simulations in this section at a single AGN disk sound speed, $c_s=10~\rm{km~s^{-1}}$ and assume that outcomes of stellar evolution such as final masses and rate of mass loss scale with $\rho c_s$.
This is not completely right, since changes in $c_s$ \emph{do} affect the stellar atmosphere, and the accretion rate scales as $\rho c_s^{-3}$ before accretion is limited by tides, which can have significant effects on evolution of stars while they are at lower masses. Thus, care must be taken when extrapolating these results to other sound speeds.

\begin{figure}
\centering
\includegraphics[width=.992\columnwidth]{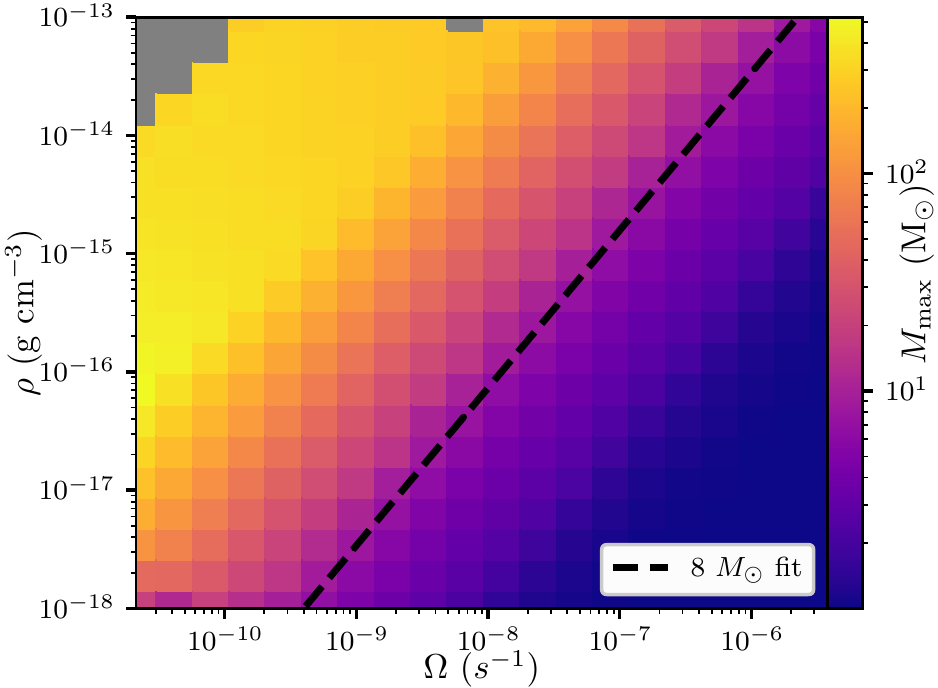}
\caption{The maximum mass achieved by each model is shown as a function of $\rho$ and $\Omega$. Gray squares indicate models that failed during boundary condition relaxation. The dashed black line is the power-law fit to $M_{\rm{max}}=8\,\Msun$ over this range, given by Equation \ref{m8equation}. Note that all runs are calculated with $c_s=10~\rm{km~s^{-1}}$.}
\label{mass2d}
\end{figure}

\begin{figure}
\centering
\includegraphics[width=.992\columnwidth]{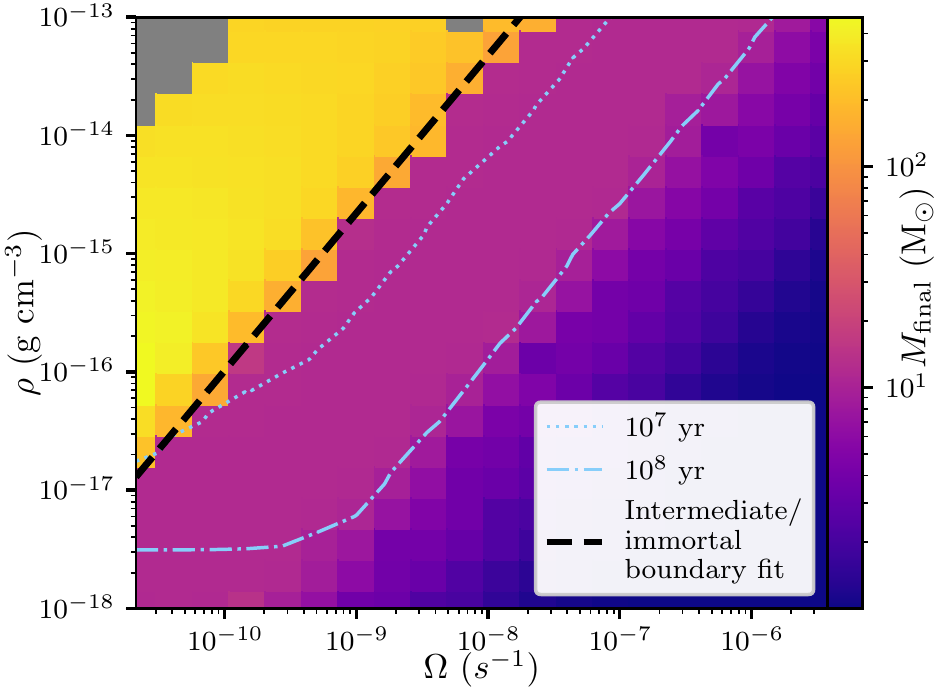}
\caption{The final mass achieved by each model is shown as a function of $\rho$ and $\Omega$. Gray squares indicate models that failed during boundary condition relaxation. The dashed black line is the power-law fit given by Equation \ref{immortalEquation} for the intermediate-immortal transition in $\rho$ and $\Omega$. The dashed thin blue line indicates models with final ages of $10^7$ years, and the dot-dashed thin blue line indicates models with final ages of $10^8$ years. Here, the reported age does not include time over which the boundary conditions were relaxed for each model. Note that all runs are calculated with $c_s=10~\rm{km~s^{-1}}$.}
\label{mass2dfin}
\end{figure}

\begin{figure*}
\centering
\includegraphics[]{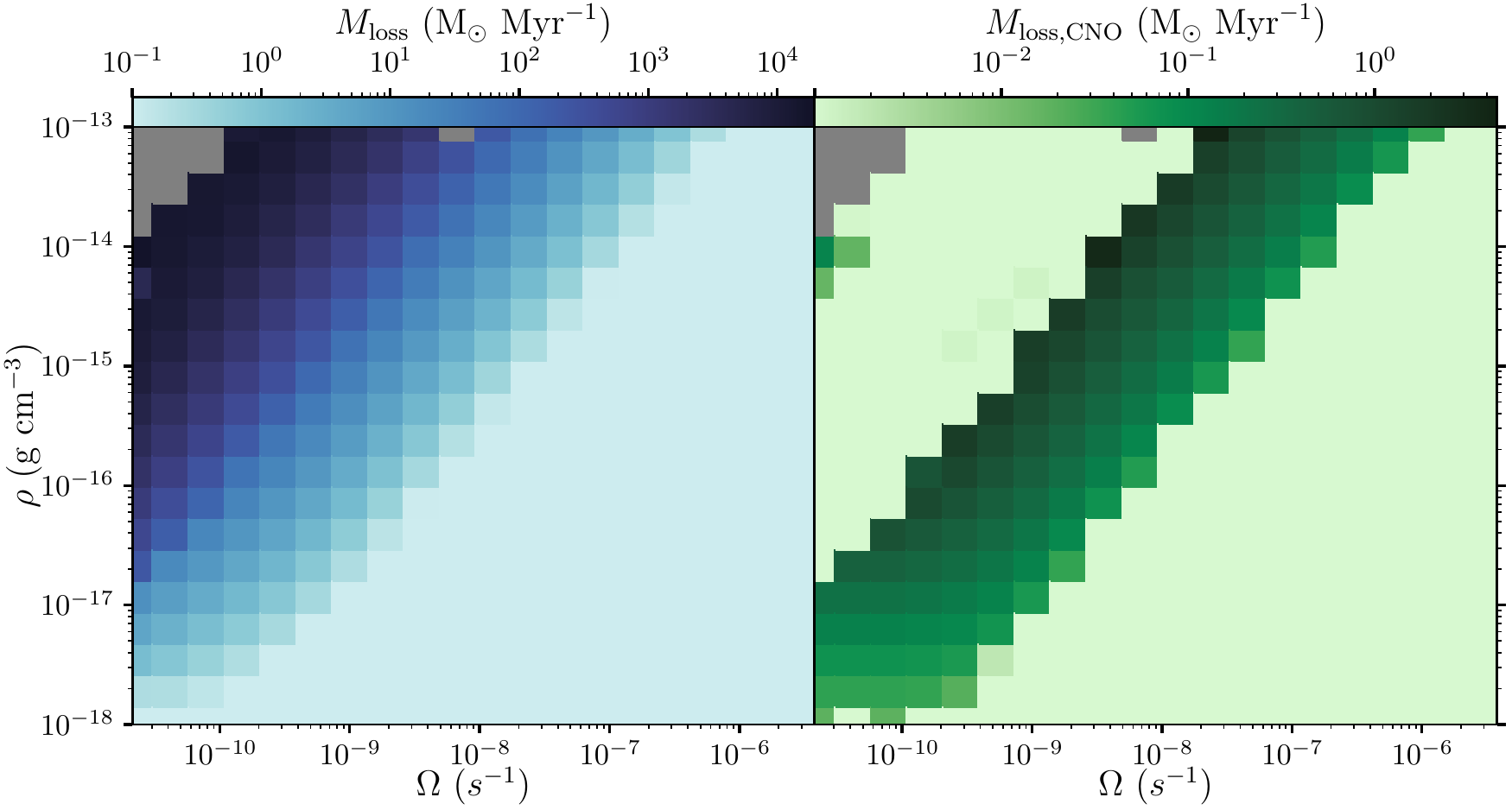}
\caption{Average rates of mass (left: total, right: carbon, nitrogen, oxygen) loss from each model as a function of $\rho$ and $\Omega$. Gray squares indicate models that failed during boundary condition relaxation. The three models at high $\rho$ and low $\Omega$ with unusually large CNO mass loss failed after boundary condition relaxation and significant accretion, but before much chemically homogeneous evolution. Note that all runs are calculated with $c_s=10~\rm{km~s^{-1}}$}
\label{yields2d}
\end{figure*}

One of the most important features of the evolution of AGN stars is that initially low-mass stars can accrete enough gas from the disk and become massive stars, ending their lives as compact objects. Additionally, since these stars reach the Eddington luminosity, they also lose large amounts of processed material before reaching the end of their evolution. Thus, in Figure~\ref{mass2d} we present the maximum masses achieved by AGN stars in our models. 

To a good degree of accuracy, quantities such as maximum and final masses can be determined based on the mass-independent factors that control the accretion rate onto AGN stars. For sufficiently large $\Omega$, such that tidal effects are significant, $\dot{M}\propto \rho\,c_s\, \Omega^{-4/3}$. Thus, a corresponding power-law contour in the $\rho-\Omega$ plane is given by $\rho=A\,c_s^{-1}\,\Omega^{4/3}$ for some constant $A$.
One such curve for $M_{\rm max}=8\,\Msun$ is marked by a black dashed line in Figure \ref{mass2d} and is given by
\begin{equation}\label{m8equation}
\left(\frac{\rho}{\rm{g~cm^{-3}}}\right) \gtrsim 
3.4\times10^{-6}\left(\frac{c_s}{\rm{10^6~cm~s^{-1}}}\right)^{-1}\left(\frac{\Omega}{\rm{s^{-1}}}\right)^{4/3}.
\end{equation}
In the low-$\Omega$ limit, as seen in Figure \ref{manypresc}, quantities such as the maximum and final mass over the course of an AGN star's life become independent of $\Omega$, and $\rho\gtrsim 5\times10^{-19}~\rm{g}~\rm{cm^{-3}}$ at $c_s=10^6~\rm{cm~s^{-1}}$ is sufficient for stars to reach greater than $8\,\Msun$.
Thus, Equation (\ref{m8equation}) can be used to predict in which regions of a disk stars will become massive before the end of their lives, at which point they will form compact objects. This result may be useful for  studies of gravitational waves involving AGN disks, as well as interpretations of anomalous AGN flares.

The disk conditions where stars become `immortal' can be seen in the high-$\rho$ low-$\Omega$ area of Figure \ref{mass2d} where maximum masses begin to plateau. However, this distinction is easier to see in terms of the final mass of a star, or in the case of stars in the immortal regime, their asymptotic stellar mass. This quantity is plotted in Figure \ref{mass2dfin}, which includes a power-law fit to the intermediate-immortal boundary, given by 
\begin{equation} \label{immortalEquation}
\left(\frac{\rho}{\rm{g~cm^{-3}}}\right) \gtrsim 
2.2\times10^{-3}\left(\frac{c_s}{\rm{10^6~cm~s^{-1}}}\right)^{-1}\left(\frac{\Omega}{\rm{s^{-1}}}\right)^{4/3}.
\end{equation}
In the low-$\Omega$ limit, one can estimate based on Figure \ref{manypresc}
that $\rho \gtrsim 2\times10^{-18} - 1\times10^{-16}\rm{~g~cm^{-3}}$ at $c_s=10^6~\rm{cm~s^{-1}}$ gives the location of the boundary, depending on ones assumption about radiative feedback.
This expression may be useful for predicting the upper(lower) limits on $\rho$($\Omega$) where AGN stars may undergo supernova or other transient events at the end of their lives. We also include contours for stellar ages in Figure \ref{mass2dfin}, which can be used to estimate whether or not a given AGN star would be able to reach its final mass within a disk lifetime. For disk lifetimes of $\sim10-100$ Myr, most intermediate stars will have time to reach the final stages of their evolution, even if initailly only $1~\rm{M_\odot}$.  Additionally, as shown in Figure \ref{exampleYields}, both the rate of stellar mass loss and its composition can vary significantly for immortal stars. 

We further investigate the mass loss rates of AGN stars and the metal content of the lost mass. From the left panel of Figure \ref{yields2d}, we see that at low $\rho/c_s^{3}$ or high $\Omega$ the total mass lost from a model is insignificant, while the opposite is true for stars that accrete more rapidly. This may be useful for gauging the kinematic impact of winds from AGN stars on accretion disks. For example, in the disk models of \citet{2005ApJ...630..167T}, there is a contribution to the pressure support from stellar feedback that is independent of disk opacity, there attributed to supernovae. However, the extreme mass loss rates from immortal stars could also provide significant pressure support and alter the disk commensurately. For example, consider an immortal star with $M_*\sim400~\Msun,$ losing mass at a rate of $\sim 10^3~\rm{M_\odot/Myr}$. We find that typical immortal stars have escape velocities in excess of $10^8~\rm{cm~s^{-1}}$, not unlike massive OB stars \citep{Lamers:1999,2014ARA&A..52..487S}. Using the escape velocity as a rough estimate of the outflow speed, this gives a ram pressure at the Hill radius ($R_H\sim10^{16}~\rm{cm}$ for $M_*\sim400\,\Msun$) of $\sim10^{-2}~\rm{erg~cm^{-3}}$. Depending on the accretion disk conditions, this can be well in excess of the ambient pressure of the AGN disk, $\rho c_s^2$, which ranges from $10^{-6}$ to $10^{-1}~\rm{erg~cm^{-3}}$ for the disk conditions explored in this work. Depending on the number of AGN stars within a given accretion disk, they may provide a significant fraction of the total pressure support. 

Only the ejecta from intermediate stars is particularly metal rich, since those stars are able to progress through later stages of evolution, as demonstrated in the right panel of Figure \ref{yields2d}. On the other hand, immortal stars have ejecta that is overall lower in metals, but is relatively more nitrogen rich (Figure~\ref{exampleYields}), as their energy generation is dominated by the CNO cycle. Stars in the intermediate regime may also lose a significant amount of metal-enriched material during a supernova, but we have not included this contribution in our current analysis.

\begin{figure*}[!ht]
\centering
\includegraphics[width=.992\textwidth]{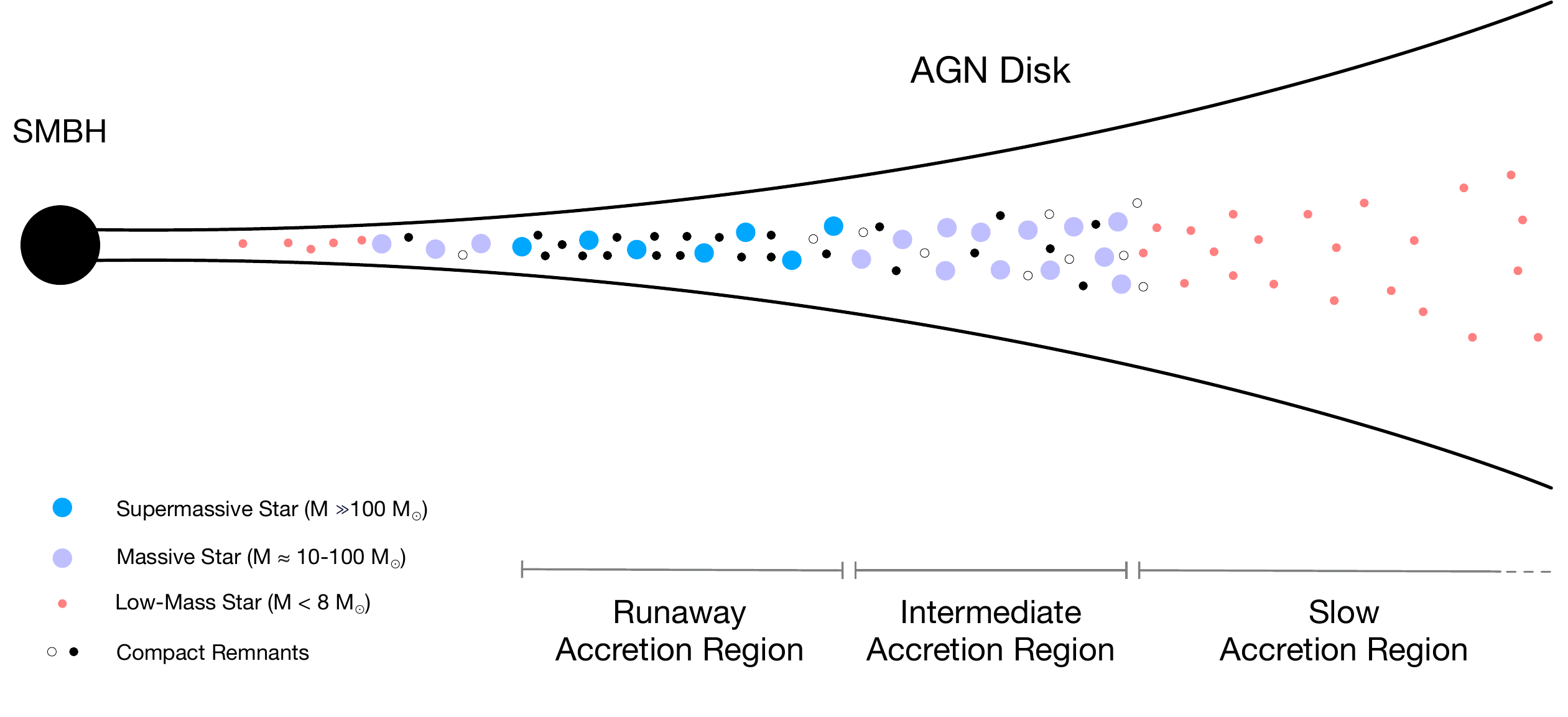}
\caption{Illustration of different regimes of accretion in an AGN disk. Stellar seeds are expected to become massive and supermassive in the intermediate and runaway accretion region, respectively. These stars pollute the disk with their super-Eddington winds and stellar explosions (SNe and GRBs). They also leave behind compact remnants that can dynamically interact and merge. In the slow accretion region, stars only gain modest amounts of material. These stars are expected to outlive the AGN phase. Note that the location of these different regions depends on the specific disk model. The accretion rate decreases  towards the SMBH because of the increasing importance of tidal effects and, to a smaller extent, shear.  The succession shown in this illustration assumes a constant sound speed through the disk and a monotonic density increase
towards the super massive black hole (SMBH). Realistic disk models might lead to different ordering. }
\label{illustration}
\end{figure*}

\section{Astrophysical Implications}
The results of stellar evolution in AGN disks depend strongly not only on the ambient gas density and temperature, but also the strength of the gravity of the SMBH on the surrounding gas. This can lead to a complex radial dependence on the effects of stars on a given accretion disk. One such example is sketched in Figure \ref{illustration}, which illustrates how at large distances accretion may be slow and stellar evolution normal due to the low gas density, but closer to the SMBH accretion may be staved off by the gravity of the black hole. Between these extremes, AGN stars may result in a number of astrophysical phenomena, from explosive transients to enriching the metallicity of the AGN disk. 

Massive stars in AGN disks, and the near-Eddington mass loss they experience, can alter the accretion disk composition and structure significantly. As shown in Figure \ref{yields2d}, AGN stars in the intermediate accretion regime can supply the disk with metals at rate of $\sim0.1-2~M_\odot$ per Myr, even when accreting a mixture containing only hydrogen and helium. Additionally, as shown in Figure \ref{mass2dfin}, many of these stars end their lives with masses $\gtrsim 10~\Msun$, and may further enrich the disk and surrounding regions, as suggested by \citet{1993ApJ...409..592A}. Through both mass loss and supernovae, metallicity enrichment has a sensitive dependence on the disk properties $\rho,~c_s$, and $\Omega$. Metallicity enrichment could be further complicated by migration through the disk, \citep[e.g.][]{{2002ApJ...565.1257T},{2010MNRAS.401.1950P},{2010ApJ...715L..68L}}, which can also be altered by the near-Eddington luminosities of AGN stars \citep[e.g.][]{{2020ApJ...902...50H}}. Detailed modeling along those lines is beyond the scope of the present work, but Equations (\ref{m8equation}) and (\ref{immortalEquation}) may prove useful to such efforts.

Despite these difficulties, AGN stars are a promising channel for producing the supersolar metallicities commonly observed in both high- and low-redshift AGN \citep[e.g.][]{{1989ApJ...347..195S},{2002ApJ...564..592H},{2019A&ARv..27....3M}}. Notably, AGN metallicities have been measured to be larger than that of their host galaxies, roughly independent of redshift, but correlated with SMBH mass \citep{2018MNRAS.480..345X}.
Additionally, if these stars undergo supernova explosions, they could significantly enhance the iron abundance in the disk, and could be partially responsible for the large iron abundances inferred from X-ray emission line analyses \citep[e.g.][]{{1995Natur.375..659T},{1997ApJ...488L..91N}}. 

Although the mass lost from immortal AGN stars is not as metal-rich as that from their intermediate counterparts, nitrogen makes up a much larger fraction of the metal content of their winds, as shown in Figure \ref{exampleYields}. A large population of such stars may lead to an overabundance of nitrogen with comparatively little metallicity enhancement. This scenario bears some resemblance to nitrogen-rich quasars, a subset ($\sim 1\%$) of quasars that show anomalously strong nitrogen lines and elevated nitrogen-to-carbon abundance ratios \citep[e.g.][]{{2004AJ....128..561B},{2004AJ....127..576B},{2008ApJ...679..962J},{2014MNRAS.439..771B},{2009A&A...503..721M}}. In a subset of these quasars, variability in nitrogen-to-carbon line ratios has been demonstrated on timescales of $\sim$years, sometimes attributed to tidal disruptions of evolved stars \citep{{2016MNRAS.458..127K},{2018ApJ...859....8L}}.
The portion of the population not displaying variability may be linked to immortal stars within the accretion disk. 

As shown in Figure \ref{mass2dfin}, there is a window in $\rho$ and $\Omega$ where AGN stars are both in the intermediate accretion regime, and can potentially reach the end of their lives within an AGN disk lifetime. 
For stars that primarily undergo Bondi accretion ($R_B\ll R_H$ for most of the star's growth) $\dot{M}\propto M_*^2$ and the accretion timescale is inversely proportional to the star's mass. On the other hand, for stars in comparatively high-$\Omega$ disks where accretion is limited by tidal effects ($R_B\gg R_H$), $\dot{M}\propto M^{2/3}$ and the accretion timescale grows slowly with stellar mass. In these limits one can extrapolate the contours in Figure \ref{mass2dfin} accordingly, but intermediate cases ($R_B\sim R_H$) are less straightforward.

We did not evolve our models far enough to accurately predict compactnesses, and thus stellar fates \citep{2013ApJ...762..126O,2014ApJ...783...10S}, although the preliminary core compactness calculations of \citet{2020arXiv200903936C} suggest that AGN stars might preferentially form black holes.
 Regardless, these massive stars are expected to leave behind neutron stars and stellar-mass black holes within the disk, and can potentially produce electromagnetic signatures. Supernovae may transport significant angular momentum through the disk \citep{2021ApJ...906...15M}, and are a possible source of transients observed in optical time-domain surveys \citep{2017MNRAS.470.4112G,2020arXiv201008554F}. Many AGN stars are expected to reach the end of their lives rapidly rotating due to the angular momentum gained as they accrete, so black holes formed in the collapse of AGN stars could generate gamma-ray bursts, which could in turn take on a number of appearances depending on their location in the accretion disk \citep{Adam, 2021ApJ...906L...7P}. 

Compact objects remaining in the disk migrate due to various torques \citep{2002ApJ...565.1257T, 2010MNRAS.401.1950P,2010ApJ...715L..68L, 2020ApJ...902...50H}, a number of which depend quadratically on the mass of the object mass. Thus, heavier objects can migrate faster, leading to mergers within the disk, which could make up a large fraction of LIGO sources \citep[e.g.][]{2017MNRAS.464..946S,2017ApJ...835..165B,2020MNRAS.498.4088M}. These objects may also migrate inward through the disk, eventually merging with the SMBH, leading to extreme- or intermediate-mass inspirals detectable by LISA \citep{2011PhRvL.107q1103Y,2019MNRAS.486.2754D,2021MNRAS.501.3540D}, and facilitating SMBH growth without being subject to the Eddington limit \citep{2020MNRAS.493.3732D}.

It is believed that the Milky Way experienced an AGN phase about 2-8 Myr ago \citep{Su:2010,Bland-Hawthorn:2019}.
With its directly available observations of stellar populations and stellar remnants, the Galactic Center (GC) is a prime target for testing the possible impact of AGN stars evolution \citep{2020arXiv200903936C}.
The central parsec contains an unexpected large number of young massive stars \citep{Ghez:2003,Alexander:2005}, with O/WR stars confined to the inner 0.5~pc region \citep{Paumard:2006,Bartko:2010}. The present day mass function of these stars is top-heavy \citep{Genzel:2010}. Contrary to theoretical expectations \citep{Bahcall:1976,Bahcall:1977}, the relative fraction of low-mass stars decreases moving towards the GC \citep{Genzel:2010,Do:2017}. Spectroscopy reveals that some of the stars in the GC may be He-rich \citep{Martins:2008,Habibi:2017,Do:2018}. Finally, \citet{Hailey:2018} found that low-mass X-ray binary candidates appear to be segregated to the inner 1~pc.

In the context of AGN stars evolution, low-mass stars are expected to become massive via accretion in the inner, denser regions of an AGN disk \citep{2020MNRAS.498.3452D}. Stars surviving the AGN phase might still carry the signature of a disk chemically enriched via AGN star evolution. Compact remnants are also expected to be radially segregated, although migration effects might play an important role as well \citep{2020ApJ...898...25T}.
While there are a number of possible competing explanations, it is intriguing that AGN stars evolution could simultaneously account for these puzzling observations of the GC.

\section{Conclusions}
Stellar evolution in AGN disks primarily depends on the properties of the disk around the star and the balance between the gravity of the star and that of the SMBH. Uncertainty in the effects of radiative feedback on the accretion stream only significantly affects our models far from the SMBH ($\Omega \lesssim 10^{-10}~\rm{s^{-1}}$, $r_\bullet \gtrsim 0.36~\rm{pc}$ for a $10^8~\Msun$ black hole). Additionally, other effects that can reduce accretion onto AGN stars are less significant than tidal forces from the SMBH. Therefore, accounting for tidal forces alone is likely sufficient to study the evolution of stars throughout AGN disks. 

Accordingly, we performed a survey in $\rho/c_s^3, ~\Omega$ to map these key parameters to the outcomes of stellar evolution in AGN disks. All of our models were initially $1~\Msun$, although a different initial mass would simply change the time required for each star to accrete significantly ($M_*/\dot{M}\propto1/M_*$). Using this survey, we have determined mappings between $\rho/c_s^3$ and $\Omega$ that can be used to predict, given an accretion disk model, whether explosive transients and compact remnants are expected (Figure \ref{mass2d}, Equation(\ref{m8equation})) or whether stars are expected to becomes super-massive ($M_*>100~M_\odot$) and live indefinitely (Figure \ref{mass2dfin}, Equation (\ref{immortalEquation})). Depending on the structure of a given accretion disk and the surrounding stellar population, a variety of outcomes are possible, such as that illustrated by Figure \ref{illustration}. AGN stars can, for example, enhance the metallicity of the accretion disk, lead to mergers of objects within the disk and accompanying gravitational waves, cause luminous transients, lend significant pressure support to the accretion disk, and leave behind stellar population with a top-heavy mass function after the dispersal of the disk. 

\section*{Software}
\texttt{MESA} \citep[][\url{http://mesa.sourceforge.net}]{{2011ApJS..192....3P},{2013ApJS..208....4P},{2015ApJS..220...15P},{2018ApJS..234...34P},{2019ApJS..243...10P}},
\texttt{MESASDK} \citep{mesasdk_linux},
\texttt{matplotlib} \citep{4160265}, \texttt{numpy} \citep{5725236}

\section*{Acknowledgments}

We are grateful to Sivan Ginzburg and Cole Miller for useful comments and discussions, and the anonymous referee whose comments helped clarify and improve our work. The Center for Computational Astrophysics at the Flatiron Institute is supported by the Simons Foundation. Computations were performed using the Rusty cluster of the Flatiron Institute and the YORP cluster administered by the Center for Theory and Computation within the Department of Astronomy at the University of Maryland.

\appendix

\section{Software details}\label{apndx:A}
Calculations were carried out using MESA version 15140. The MESA EOS is a blend of the OPAL \citep{Rogers2002}, SCVH
\citep{Saumon1995}, FreeEOS \citep{Irwin2004}, HELM \citep{Timmes2000},
and PC \citep{Potekhin2010} EOSes.

Radiative opacities are primarily from OPAL \citep{Iglesias1993,
Iglesias1996}, with low-temperature data from \citet{Ferguson2005}
and the high-temperature, Compton-scattering dominated regime by
\citet{Buchler1976}.  Electron conduction opacities are from
\citet{Cassisi2007}.

Nuclear reaction rates are from JINA REACLIB \citep{Cyburt2010} plus
additional tabulated weak reaction rates \citet{Fuller1985, Oda1994,
Langanke2000}.  
Screening is included via the prescription of \citet{Chugunov2007}.
Thermal neutrino loss rates are from \citet{Itoh1996}.

We adopted a 21-isotope nuclear network (approx21.net). We used the Schwarzschild criterion to determine convective boundaries and did not include convective overshooting. Inlists and source code used in this paper can be found at \url{https://zenodo.org/record/5071458}.

\bibliographystyle{aasjournal}
\bibliography{references}

\end{document}